\documentclass[%
superscriptaddress,
twocolumn,
prab,
]{revtex4-1}

\usepackage{graphicx}% Include figure files
\usepackage{dcolumn}% Align table columns on decimal point
\usepackage{bm}% bold math
\usepackage{amsmath,amssymb}
\usepackage{siunitx}
\usepackage{color}
\usepackage[english]{babel}
\usepackage{amsfonts}
\begin{document}

\preprint{APS/123-QED}

\title{Meter-Scale, Conditioned Hydrodynamic Optical-Field-Ionized Plasma Channels}

\author{A. Picksley}%
\affiliation{John Adams Institute for Accelerator Science and Department of Physics,University of Oxford, Denys Wilkinson Building, Keble Road, Oxford OX1 3RH, United Kingdom}%
\author{A. Alejo}%
\affiliation{John Adams Institute for Accelerator Science and Department of Physics,University of Oxford, Denys Wilkinson Building, Keble Road, Oxford OX1 3RH, United Kingdom}%
\author{R. J. Shalloo}%
\affiliation{John Adams Institute for Accelerator Science and Department of Physics,University of Oxford, Denys Wilkinson Building, Keble Road, Oxford OX1 3RH, United Kingdom}%
\author{C. Arran}
\affiliation{John Adams Institute for Accelerator Science and Department of Physics,University of Oxford, Denys Wilkinson Building, Keble Road, Oxford OX1 3RH, United Kingdom}%
\author{A. von Boetticher}
\affiliation{John Adams Institute for Accelerator Science and Department of Physics,University of Oxford, Denys Wilkinson Building, Keble Road, Oxford OX1 3RH, United Kingdom}%
\author{L. Corner}%
\affiliation{Cockcroft Institute for Accelerator Science and Technology, School of Engineering, The Quadrangle, University of Liverpool, Brownlow Hill, Liverpool L69 3GH, United Kingdom}%
\author{J. A. Holloway}%
\affiliation{John Adams Institute for Accelerator Science and Department of Physics,University of Oxford, Denys Wilkinson Building, Keble Road, Oxford OX1 3RH, United Kingdom}%
\author{J. Jonnerby}%
\affiliation{John Adams Institute for Accelerator Science and Department of Physics,University of Oxford, Denys Wilkinson Building, Keble Road, Oxford OX1 3RH, United Kingdom}%
\author{O. Jakobsson}%
\affiliation{John Adams Institute for Accelerator Science and Department of Physics,University of Oxford, Denys Wilkinson Building, Keble Road, Oxford OX1 3RH, United Kingdom}%
\author{C. Thornton}%
\affiliation{Central Laser Facility, STFC Rutherford Appleton Laboratory, Didcot OX11 0QX, United Kingdom}%
\author{R. Walczak}%
\affiliation{John Adams Institute for Accelerator Science and Department of Physics,University of Oxford, Denys Wilkinson Building, Keble Road, Oxford OX1 3RH, United Kingdom}%
\author{S. M. Hooker} \email{simon.hooker@physics.ox.ac.uk}
\affiliation{John Adams Institute for Accelerator Science and Department of Physics,University of Oxford, Denys Wilkinson Building, Keble Road, Oxford OX1 3RH, United Kingdom}%

\date{\today}

\begin{abstract}
We demonstrate through experiments and numerical simulations that low-density, low-loss, meter-scale plasma channels can be generated by employing a conditioning laser pulse to ionize the neutral gas collar surrounding a hydrodynamic optical-field-ionized (HOFI) plasma channel.  We use particle-in-cell simulations to show that the leading edge of the conditioning pulse ionizes the neutral gas collar to generate a deep, low-loss plasma channel which guides the bulk of the conditioning pulse itself as well as any subsequently injected pulses. In proof-of-principle experiments we generate conditioned HOFI (CHOFI) waveguides with axial electron densities of $n_\mathrm{e0} \approx 1 \times 10^{17} \; \mathrm{cm^{-3}}$, and a matched spot size of $26 \; \mathrm{\mu m}$. The power attenuation length of these CHOFI channels was calculated to be $L_\mathrm{att} = (21 \pm 3) \; \mathrm{m}$, more than two orders of magnitude longer than achieved by HOFI channels.  Hydrodynamic and particle-in-cell simulations demonstrate that meter-scale CHOFI waveguides with attenuation lengths exceeding 1 m could be generated with a total laser pulse energy of only $1.2$ J per meter of channel. The properties of CHOFI channels are ideally suited to many applications in high-intensity light-matter interactions, including multi-GeV plasma accelerator stages operating at high pulse repetition rates.

This article was published in Physical Review E \textbf{102}, 053201 on 2 November
2020. DOI: 10.1103/PhysRevE.102.053201

\textcopyright  2020 American Physical Society.

\end{abstract}

\maketitle

\section{Introduction}

Many fields exploiting laser-plasma interactions require the laser pulse to propagate through plasma for many Rayleigh ranges, which means that the laser pulse must be guided via relativistic self-focusing effects \cite{Sprangle:1987, Sun1987, Sprangle:1990a} or by an external waveguide. Examples of such applications include Raman amplification \cite{Trines:2010ir}, high-harmonic generation in ions, x-ray lasers \cite{Butler:2003}, and laser-driven plasma accelerators \cite{Tajima1979, Esarey:2009, Hooker2013jk}. The last of these is particularly challenging since the multi-GeV accelerator stages needed to drive compact light sources, or future particle colliders \cite{Walker:2017hi,USroadmap}, require guiding of laser pulses with peak intensities of order $\SI{E18}{W.cm^{-2}}$ over distances in the range \SIrange{0.1}{1}{m} through plasma with a density of order $\SI{E17}{cm^{-3}}$.

Many methods for guiding high-intensity laser pulses have been investigated. These include grazing-incidence guiding in capillaries \cite{Cros:2002}, and many varieties of plasma channels generated by hydrodynamic expansion \cite{Durfee:1993, Durfee:1995gr, Sheng2005, Smartsev:2019cy}, capillary discharges \cite{Ehrlich:1996, Spence:2000fr, Butler:2002zza}, Z-pinches \cite{Hosokai:2000, Luther:2005eu}, open-geometry discharges \cite{Lopes:2003}, and laser-heated capillary discharges \cite{Bobrova:2013jz,Gonsalves:2020bg}. To date, the most successful approaches for driving laser-plasma accelerators are capillary discharges and its laser-heated variant, which have been used to generate electron beams with energies up to \SI{4.2}{GeV} and \SI{7.8}{GeV} respectively \cite{Leemans:2014kp,Gonsalves:2019ht}. However, a remaining challenge for this approach is avoiding laser damage to the capillary structure, which would limit the working lifetime of the waveguide in future high repetition rate plasma accelerators.

Hydrodynamic plasma channels \cite{Durfee:1993, Durfee:1995gr} have the major advantage of being free-standing, and hence immune to laser damage. In this approach a column of plasma is created, and heated collisionally, by a picosecond-duration laser pulse. Rapid radial expansion of the plasma column drives a strong shock wave into the surrounding gas, forming a plasma channel --- i.e. a region of radially-increasing electron density --- between the axis and the shock front. In order to achieve sufficiently rapid heating the initial density must be high, which limits the on-axis density of the subsequent channel to $n_\mathrm{e0} \gtrsim \SI{1E18}{cm^{-3}}$.

We recently proposed \cite{Shalloo:2018fy} that optical field ionization could generate much lower density channels since the temperature to which it heats the ionized electrons is independent of plasma density. In previous work \cite{Shalloo:2019hv, Picksley:2020} we have generated these hydrodynamic optical-field-ionized (HOFI) plasma channels with lengths as long as \SI{100}{mm},  with axial densities as low as $n_\mathrm{e0} \lesssim \SI{1e17}{cm^{-3}}$, and at repetition rates of up to \SI{5}{Hz}. 

A drawback of HOFI channels is that they are shallower than collisionally-heated hydrodynamic channels, since the energy per unit length deposited in the initial plasma column is nearly an order of magnitude lower  \cite{Clark:1997we}, and since the channel depth is roughly proportional to the axial density \cite{Shalloo:2019hv}. Here channel depth is defined as the difference between the peak electron density which occurs at the shock front and the axial electron density, $\Delta n_\mathrm{e} = n_\mathrm{e}(r_\mathrm{shock}) - n_\mathrm{e0}$. As a consequence, the power attenuation lengths achieved to date in HOFI channels \cite{Picksley:2020} are of order $L_\mathrm{att} \lesssim \SI{100}{mm}$, which is too short for many applications.

In the present paper we build on a phenomenon observed in our 2018 experiments on HOFI channel formation with axicon lenses \cite{Shalloo:2019hv}: the transverse extent of the plasma immediately after the passage of the guided pulse is greater than before that pulse arrives. This suggests that the neutral species surrounding the plasma column are ionized by the electric fields of the guided pulse leaking through the channel walls \cite{robert2018a}. In this paper we present that data in detail. We show, through experiments and simulations, that guiding a ``conditioning pulse'' in a HOFI channel can create a  deep, thick-walled plasma channel by ionization of the collar of neutral gas which surrounds the initial HOFI channel. We present measurements showing the evolution of the transverse electron density profile of these conditioned HOFI (CHOFI) channels, and demonstrate the formation of \SI{16}{mm}-long CHOFI channels at a repetition rate of \SI{5}{Hz} (limited only by the repetition rate of the laser system), with power attenuation lengths of  $\sim \SI{20}{m}$.  We use hydrodynamic and particle-in-cell (PIC) simulations to understand the formation of the CHOFI channels, and to demonstrate that they could be generated at the metre scale with modest laser pulse energies.

We note that this work is closely related to earlier work by Spence and Hooker \cite{Spence:2000fr}, who investigated `quasi-matched' guiding in which ionization of a partially-ionized parabolic plasma channel by the leading edge of an intense laser pulse creates a fully-ionized plasma channel which can guide the bulk of the pulse with low losses. More recently,  Morozov et al.\ \cite{Morozov:2018cr} explored similar effects to those reported here in their investigation of ionization-assisted guiding in plasma channels formed in \SI{3.5}{mm} long gas jets with an initial atomic density of order $\SI{1E19}{cm^{-3}}$. An alternative approach in which the neutral gas collar is ionized by a coaxial high-order Bessel beam has also recently been described \cite{Miao:2020ir}.

\section{Experimental Setup}
\begin{figure}[bt]
    \centering
    \includegraphics[width=86mm]{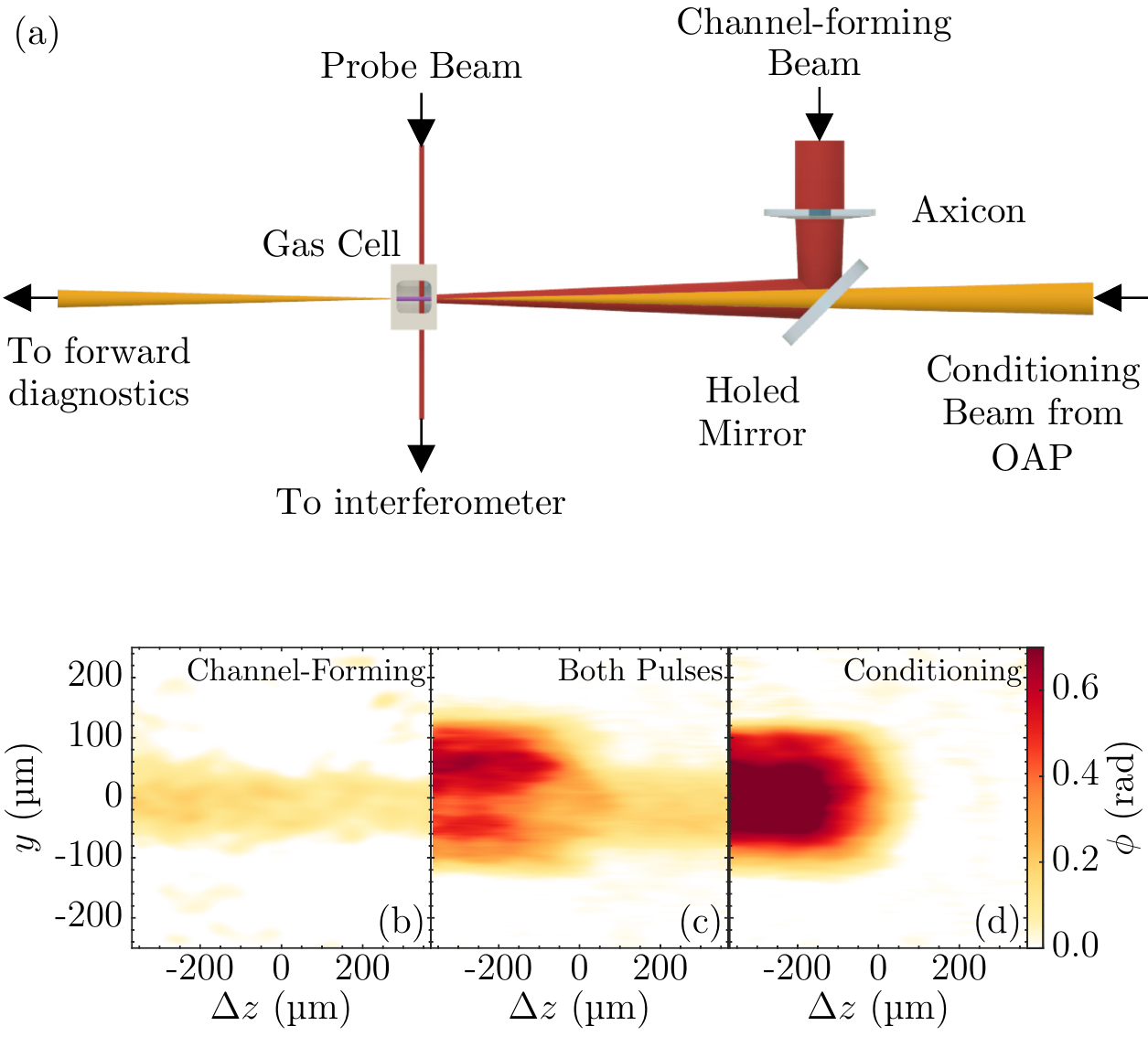}
    \caption{(a) Schematic of the experimental interaction region. (b-d) Phase shifts measured by the transverse probe beam at $z \approx \SI{3.5}{mm}$ for: (b) the channel-forming pulse alone, (c) the channel-forming pulse and the conditioning pulse at a delay $\tau = \SI{1.5}{ns}$; and (d) the conditioning pulse alone}
    \label{fig:exp_ionization}
\end{figure}

The experiments were undertaken with the \SI{5}{Hz} repetition-rate Astra-Gemini TA2 laser at the Central Laser Facilty, UK. The set-up employed has been described previously \cite{Shalloo:2019hv}, and is described in detail in the Supplemental Material \cite{supp_mat}, so here we provide only an outline.

The compressed, linearly-polarized beam from the Astra-Gemini TA2 laser was split into a channel-forming beam and a guided or ``conditioning'' beam by an annular dielectric mirror. The channel-forming beam reflected from this mirror was transformed to circular polarisation by a quarter waveplate. It was subsequently directed to a retroreflecting delay stage, focused by a fused silica axicon lens of base angle $\vartheta = \SI{5.6}{\deg}$, and directed into a gas cell by an annular turning mirror. The axicon generated a line focus, which extended throughout the length of the gas cell, with a peak intensity on axis of approximately $\SI{5e15}{W.cm^{-2}}$. 

\begin{figure*}[t!] 
    \centering
    \includegraphics[width=\textwidth]{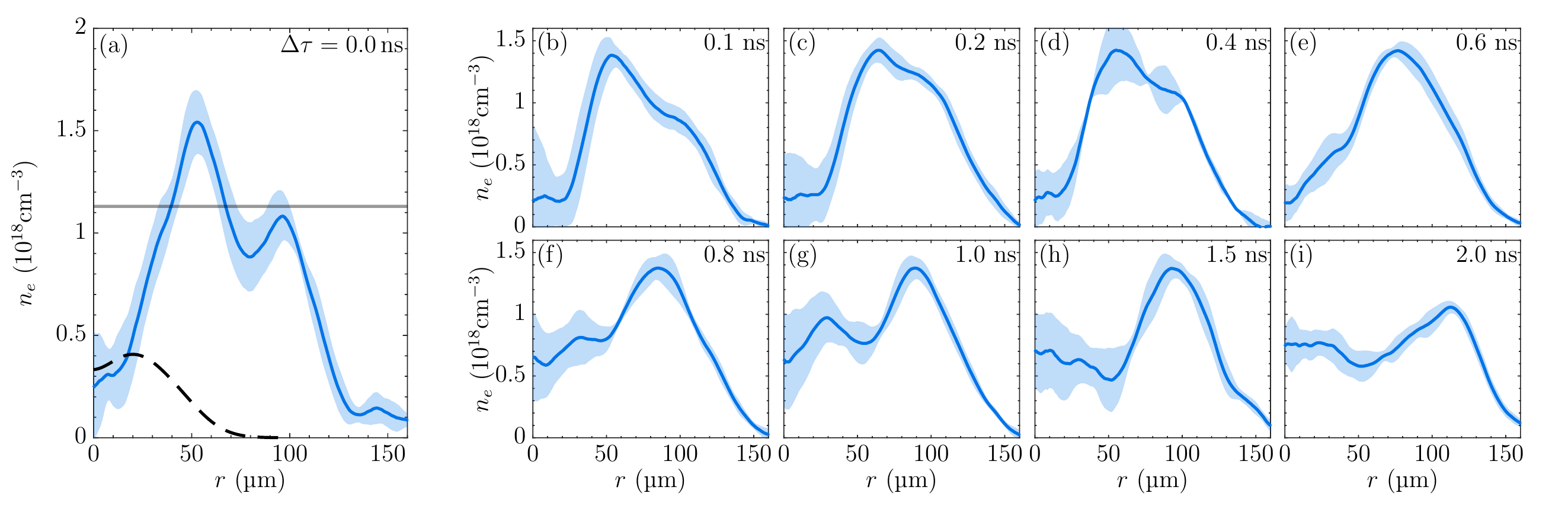}
    \vspace{-8mm}
    \caption{Formation and temporal evolution of the transverse electron density profiles $n_\mathrm{e}(r)$ in the CHOFI waveguide.  (a) Measured electron density profile immediately before (dashed) and after (solid) the arrival of the conditioning pulse at $\tau = \SI{1.5}{ns}$ ($\Delta \tau = \SI{0}{ns}$). The horizontal, grey line shows the density corresponding to full ionization of the ambient gas.  (b)-(i) Electron density profiles measured at different additional delays $\Delta \tau$ (indicated) after the arrival of the channel-forming pulse, for a conditioning pulse arriving at  $\tau = \SI{1.5}{ns}$. All plots show the mean of approximately 20 shots recorded at a repetition-rate of \SI{5}{Hz}, with the probe pulse crossing at $z \approx \SI{3.5}{mm}$ with respect to the front pinhole. The data is averaged longitudinally over approximately $\SI{0.75}{mm}$. The shaded region is one standard deviation wide. Since the measured phase shift was significantly smaller for $\Delta \tau < 0$, the dashed electron density profile shown in (a) was retrieved using a forward fitting method as described in Ref. \cite{Shalloo:2019hv}.}
    \label{fig:exp_evolution}
\end{figure*}

The conditioning pulse transmitted through the annular mirror was sent to a retro-reflecting delay stage and focused through the hole in the turning mirror by an off-axis paraboloid (OAP) mirror of focal length $f = \SI{750}{mm}$ to the entrance pinhole of the gas cell. The measured beam waist and Rayleigh range of the conditioning beam were $w_0 = \SI{22(3)}{\micro m}$ and $z_R = \SI{1.1(4)}{mm}$ respectively. The conditioning pulse could be operated at two intensities: (i) high-intensity, with a peak axial intensity at focus of $I_\mathrm{{peak}}^\mathrm{{high}} = \SI{4.1e17}{W.cm^{-2}}$; and (ii) low intensity, with a peak  intensity of $I_\mathrm{{peak}}^\mathrm{{low}} \approx \SI{1.6e16}{W.cm^{-2}}$, achieved by replacing one of the mirrors in the conditioning pulse beamline with an uncoated, optically flat wedge.

The channel-forming and conditioning pulses entered and left the gas cell via entrance and exit pinholes of diameter \SI{1.5}{mm} and \SI{800}{\micro m} respectively, spaced by \SI{16}{mm}. The cell pressure could be varied in the range \SIrange{5}{120}{mbar}.

Optically flat fused silica windows mounted on the sides of the cell allowed optical access by a separately compressed, $\SI{1}{mJ}$, \SI{800}{nm} probe pulse used for transverse interferometry of the plasma channels. The delay $\tau$ between the arrival of the channel-forming beam and the probe beam could be varied in the range \SIrange{0}{6}{ns} by a  delay stage in the probe beam line. A Keplerian telescope magnified the transmitted probe beam by a factor of 6, and a Mach-Zehnder interferometer located within the telescope generated a fringe pattern, which was recorded by a 12-bit CMOS detector located at the image plane of the telescope. The measured spatial resolution in the object plane  of this interferometer was $\SI{8.7(9)}{\micro m}$.

A mode imaging system was used to record the exit mode of the conditioning pulse, and the energy transmission of this pulse was measured by directing a fraction of the transmitted beam to a pyroelectric energy meter.

\section{Experimental Results}
\begin{figure}[bt]
    \centering
    \includegraphics[width=86.4mm]{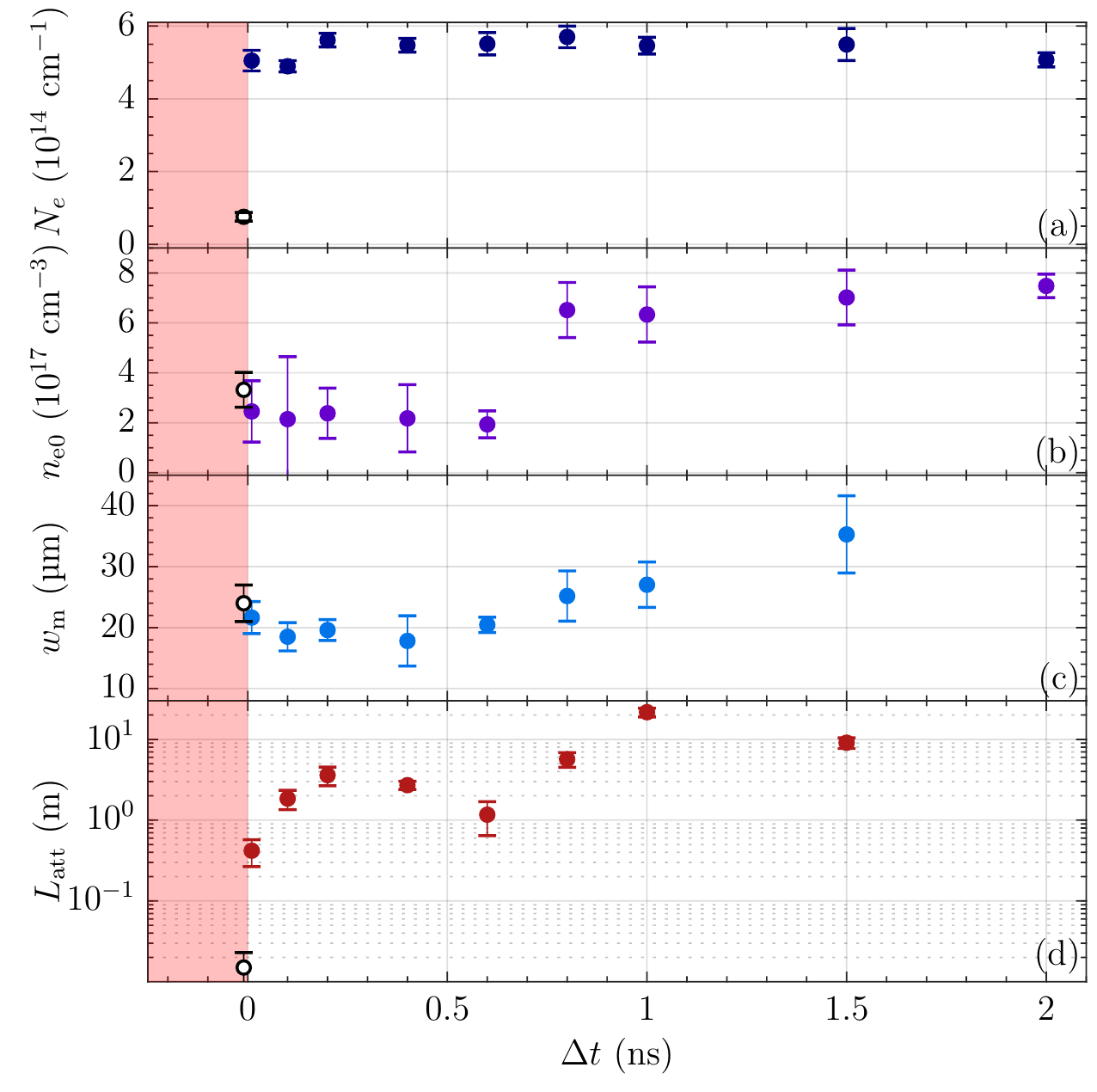}
    \caption{Temporal evolution of the properties of the CHOFI plasma channels as a function of delay $\Delta \tau$ after the arrival of the conditioning pulse:  (a) Measured total number of electrons per unit length of waveguide, $N_\mathrm{e}$; (b) the measured axial density  $n_\mathrm{e0}$; (c) the calculated matched spot size $w_\mathrm{m}$ of the lowest order mode; and (d) the calculated attenuation length of the lowest order channel mode. The white circles show the same parameters of the HOFI channel immediately before the conditioning pulse arrives, indicated by the red region. In all cases, the error bars reflect the uncertainties in $n_\mathrm{e}(r)$ depicted in Fig. \ref{fig:exp_evolution}.}
    \label{fig:exp_properties}
\end{figure}

Figure \ref{fig:exp_ionization}(b-d) captures the moment when the conditioning pulse arrives, at $\tau = \SI{1.5}{ns}$ after the channel-forming pulse. In Fig.\  \ref{fig:exp_ionization}(c) the conditioning pulse has reached the point $\Delta z = 0$, so that the region of positive $z$ contains plasma generated by the channel-forming pulse alone, whereas the region $\Delta z < 0$ contains plasma formed by both pulses. It is clear that the conditioning pulse increases the radial extent of the plasma by a factor of approximately 5 from that produced by the channel-forming pulse alone. Further, the radial extent of the plasma generated by both pulses is essentially the same as that of the plasma produced by the conditioning pulse alone, as shown in Fig.\  \ref{fig:exp_ionization}(d). It is clear from Fig.\ \ref{fig:exp_ionization} that the transverse wings of the guided conditioning pulse ionize neutral gas surrounding the HOFI plasma channel.

Figure \ref{fig:exp_evolution} shows the formation and temporal evolution of the transverse electron density profile of the CHOFI channels, obtained by Abel inversion of retrieved phase shifts like that shown in Fig.\ \ref{fig:exp_ionization}. Figure \ref{fig:exp_evolution}(a) shows the electron density profile before and immediately after the arrival of the conditioning pulse at $\tau = \SI{1.5}{ns}$. Since the phase shift measured without the conditioning pulse present was significantly smaller ($\lesssim \SI{100}{mrad}$), a forward fitting method \cite{Shalloo:2019hv} was used to retrieve $n_\mathrm{e}(r)$ for the dashed curve shown in fig. \ref{fig:exp_evolution}(a). It can be seen that the conditioning pulse has little effect on the electron density for radii within the radius of the shock front,  $r_\mathrm{shock} \approx \SI{25}{\micro m}$, created by the channel-forming pulse. However, at larger radii the electron density is increased substantially to form a deep, thick-walled CHOFI plasma channel: the depth of the  channel was increased by a factor of 10 to $\Delta n_\mathrm{e} = \SI{1.3(1)e18}{cm^{-3}}$; whereas, within the experimental uncertainty, the axial density remained at $n_\mathrm{e0} = \SI{2.4(10)e17}{cm^{-3}}$. The radial extent of the plasma was increased by the conditioning pulse from $r_\mathrm{shock} \approx \SI{25}{\micro m}$ to $r_\mathrm{max} \approx \SI{120}{\micro m}$.

As shown in Fig.\ \ref{fig:exp_evolution}, the electron density in the region $r > r_\mathrm{shock}$ is comparable to that expected for full ionization of the initial ambient gas, consistent with field ionization of the neutral gas by the transverse wings of the conditioning pulse. It is noticeable that in the region close to $r \approx \SI{50}{\micro m}$ the electron density is greater than that which would be generated by ionization of the neutral gas at its ambient density; this is consistent with a build up of neutral gas at the shock front as it pushes the gas outwards.

The evolution of the plasma at various delays $\Delta \tau$ after the arrival of the conditioning pulse at $\tau = \SI{1.5}{ns}$ is shown in Fig. \ref{fig:exp_evolution}(a)-(f). For $\Delta \tau = \SI{0.2}{ns}$, the measured axial density remained approximately the same, and the channel wall thickened slightly. At longer delays $\Delta \tau$, electrons in the high density ring spread radially outwards and inwards, which causes the position of the peak density to increase to $r = \SI{100}{\micro m}$ by $\Delta \tau \approx \SI{1}{ns}$ and the axial density to increase. Even at $\Delta \tau \approx \SI{2}{ns}$, where the channel structure has begun to decay, the measured channel depth was greater than before the arrival of the conditioning pulse.

Figure \ref{fig:exp_properties} summarizes the evolution of the properties of the CHOFI channel as a function of delay $\Delta \tau$. The evolution of the total number of electrons per unit length of waveguide $N_\mathrm{e} = 2 \pi \int^{r_\mathrm{max}}_0 n_\mathrm{e}(r) r \, \mathrm{d}r$, is shown in in Fig.\ \ref{fig:exp_properties}(a). Upon arrival of the conditioning pulse, $N_\mathrm{e}$ increased from $N_\mathrm{e} = \SI{7.6e13}{cm^{-1}}$ to $N_\mathrm{e} = \SI{5.1e14}{cm^{-1}}$ instantaneously. Following this, the number of electrons per unit length increased for the first \SI{0.2}{ns} after the conditioning pulse, indicating that the observed increase in the wall thickness observed arose from further ioniziation close to the shock front. We note that \SI{0.2}{ns} is consistent with the timescale for hot electrons to collide with neutral gas under these plasma parameters \cite{lotz1968}. Fig. \ref{fig:exp_properties}(b) shows the evolution of the axial density, showing that this remains close to that of the HOFI channel until $\Delta \tau \approx \SI{0.6}{ns}$, whereupon it increases rapidly as electrons from the channel wall move towards the axis.

The guiding properties of the CHOFI waveguide were compared to those of the HOFI channels formed by the channel-forming pulse alone. The lowest-order modes for each electron density profile shown in Fig.\ \ref{fig:exp_evolution} were calculated using the method outlined by Clark \textit{et al.} \cite{Clark:2000dk}, and the power attenuation lengths $L_{\mathrm{att}}$ of these modes were calculated by solving the paraxial Helmholtz equation \cite{Picksley:2020}. Figure \ref{fig:exp_properties}(c) shows the evolution of the matched spot size $w_\mathrm{m}$. It can be seen that $w_\mathrm{m}$ remains close to that of the HOFI channel, which was measured to be $w_\mathrm{m, HOFI} = \SI{24(3)}{\micro m}$, up to delays of $\Delta \tau \approx \SI{0.5}{ns}$, whereupon it increased  to $w_\mathrm{m} = \SI{35(6)}{\micro m}$ over the next nanosecond. The effect of the conditioning pulse on the propagation losses is striking: immediately before the arrival of the conditioning pulse, $L_{\mathrm{att}} = \SI{15(8)}{mm}$, the relatively large uncertainty arising from the small fringe shifts ($\phi \lesssim \SI{100}{mrad}$) generated by the HOFI channel \cite{Shalloo:2019hv}. Immediately after the arrival of the conditioning pulse, $L_{\mathrm{att}} = \SI{0.42(1)}{m}$.  In the following \SI{200}{ps}, the increase in $N_\mathrm{e}$ increases the thickness of the channel wall, further increasing the attenuation length. The largest $L_\mathrm{att}$ was achieved at $\Delta \tau = \SI{1.0}{ns}$, where $L_\mathrm{att} = \SI{21(3)}{m}$, almost four orders of magnitude larger than achieved by the channel-forming pulse alone.

\begin{figure}[tb]
    \centering
    \includegraphics[width=86.4mm]{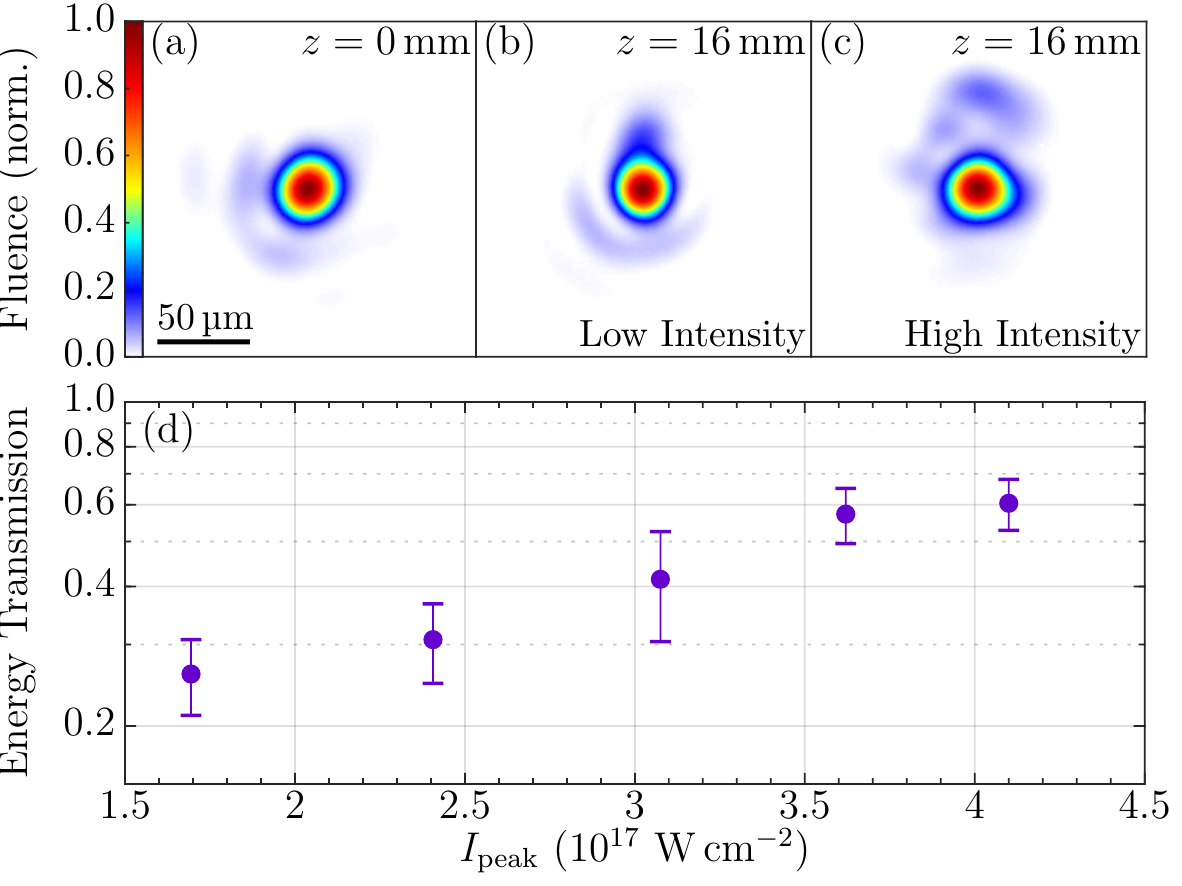}
    \caption{Properties of the transmitted conditioning pulse. The transverse fluence profile of the conditioning pulse at the input plane of the gas cell (a), and at the exit of the cell for the low- (b) and high-intensity (c) configurations. (d) the pulse energy transmission $T$ of the conditioning pulse as a function of input input $I_\mathrm{peak}$.} 
    \label{fig:exp_guided}
\end{figure}

Figure \ref{fig:exp_guided} shows the transverse fluence profile of the low- and high-intensity conditioning pulses at the entrance and exit of the \SI{16}{mm}-long gas cell for $P = \SI{40}{mbar}$ and $\tau = \SI{1.5}{ns}$. It is clear that in both cases the conditioning pulse was guided over the length of the gas cell, corresponding to approximately $14.5z_\mathrm{R}$. For the low-intensity case, the measured energy transmission and output spot radius was $T = \SI{21(4)}{\%}$ and $w_\mathrm{out} = \SI{23(4)}{\micro m}$ respectively. For the high-intensity configuration, $T = \SI{60(3)}{\%}$ and $w_\mathrm{out} = \SI{25(4)}{\micro m}$. It is clear that the mode profiles of the transmitted pulses were essentially the same for the two cases. However, the energy transmission was substantially higher for the high-intensity pulse. Figure \ref{fig:exp_guided}(d) shows the measured variation of $T$ as the peak input intensity of the pulse was increased, keeping the other parameters constant. It is clear that $T$ increases substantially as the intensity of the pulse is increased. This suggests that the leading edge of the conditioning pulse improved the channel properties via ionization, as in Fig.\ \ref{fig:exp_evolution}, allowing the body of the pulse to propagate with low losses. The coupling efficiency of the conditioning pulse, deduced from the overlap integral of the input spot with the calculated HOFI channel modes is estimated to be $T(0) \approx \SI{65}{\%}$, and hence at the highest input intensities the propagation losses of the conditioning pulse were low. It should be noted that since the channel-forming beam and conditioning beam energies could not be varied independently, the intensity of the channel-forming pulse also varied with the intensity of the conditioning pulse. However, interferometric measurements indicated that properties of the HOFI channel created by the channel-forming pulse were not significantly affected for the intensity range considered here.

\section{Hydrodynamic Simulations}
Hydrodynamic simulations were undertaken to understand in detail the distribution of plasma and neutral gas prior to the arrival of the conditioning pulse. These were performed in two dimensions, using the Eulerian code FLASH  \cite{fryxell2000flash}. A 3-temperature model was employed, which allowed for independent evolution of the electron, ion, and radiation species. The simulations included energy diffusion, thermal conductivity,  heat exchange between electrons and ions and atoms, and radiation transport. Tabulated values for the equations of state were employed. The initial conditions of the simulations were a plasma column with a super-gaussian transverse electron temperature surrounded by neutral atomic hydrogen  at a pressure of \SI{50}{mbar} and a temperature of \SI{298}{K}.

\begin{figure}[tb]
    \centering
    \includegraphics[width=\linewidth]{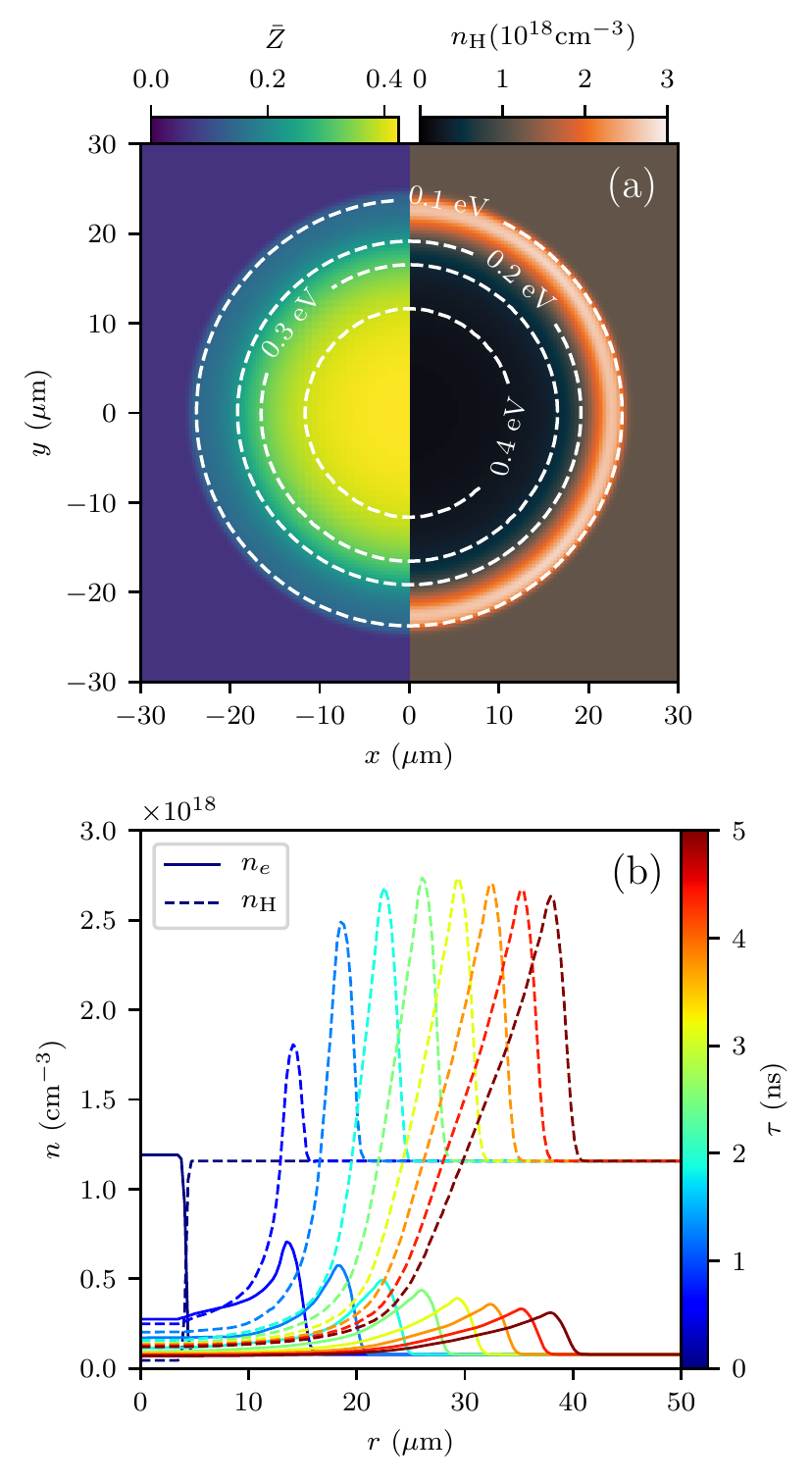}
    \caption{Hydrodynamic simulation of a HOFI channel. (a) Properties of the plasma channel at $\tau=\SI{1.5}{ns}$. In the region $x<\SI{0}{\micro m}$ is shown the ionization fraction of the hydrogen atoms ($\bar{Z}$), and in the region $x>\SI{0}{\micro m}$ the neutral hydrogen density ($n_H$).  Superimposed on these plots are contours of the electron temperatures $T_e$; the contours are spaced by \SI{0.1}{eV}, and the outer contour is at $T_e = \SI{0.1}{eV}$. (b) Temporal evolution of the electron ($n_e$, straight) and neutral hydrogen ($n_H$, dashed) transverse density profiles at various delays $\tau$ given  by the color scale.}
    \label{fig:flash_sim}
\end{figure}

Figure \ref{fig:flash_sim} summarizes the results of these simulations. Figure \ref{fig:flash_sim}(a) shows, for a delay $\tau = \SI{1.5}{ns}$, the distribution of the ionization fraction $\bar{Z}$, the density of neutral hydrogen $n_H$, and the electron temperature. The formation of a high-density shock front is clearly observed, as expected for a Sedov-Taylor-like expansion. Also evident is significant cooling of the plasma as it expands. For this delay the plasma cools on-axis from \SI{10}{eV} to \SI{0.44}{eV}, which results in the fractional ionization decreasing from $\bar{Z}\simeq \SI{100}{\%}$ to $\bar{Z}\simeq \SI{44}{\%}$. The cooling is even more pronounced at the shock front, where $T_{e,shock} \simeq 0.1-\SI{0.2}{eV}$ and hence $\bar{Z}\simeq \SI{15}{\%}$.

\begin{figure*}[tb] 
    \centering
    \includegraphics[width=\textwidth]{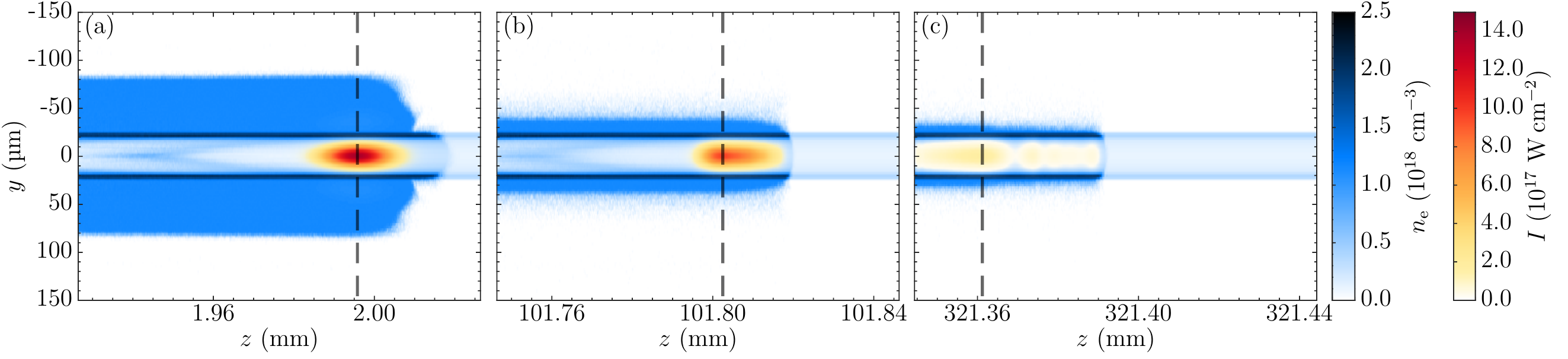}
    \caption{Transverse profiles of the electron density and the laser intensity when the peak of the conditioning pulse, indicated by the dashed line, has reached (a) $z = \SI{2}{mm}$, (b) $z = \SI{101}{mm}$, and (c) $z = \SI{321}{mm}$. The input intensity of the conditioning pulse was $I_\mathrm{peak} = \SI{6.0e17}{W.cm^{-2}}$.}
    \label{fig:sim_vis}
\end{figure*}

Figure \ref{fig:flash_sim}(b) shows the temporal evolution of the electron and neutral gas transverse density profiles. Although the initial temperature is high enough to allow for full ionization of the plasma column, a neutral gas collar appears in the early stages of the expansion. The simulation shows the number of free electrons to remain approximately constant throughout the expansion (see Supp. Material \cite{supp_mat}), indicating that the accumulation of neutral gas in the region of the shock is caused by the high pressure in the inner regions of the the channel.

\section{Propagation Simulations}
Particle-in-cell (PIC) simulations were performed in quasi-3D cylindrical geometry using FBPIC \cite{Lehe2016} to provide insight into the formation of CHOFI channels, and to explore the prospects for generating metre-scale channels. We note that use of a PIC code means that ponderomotive effects are included. The PIC simulations used the transverse electron density and neutral density profiles calculated by the FLASH simulations for a delay $\tau = \SI{1.5}{ns}$.  The conditioning laser pulse was assumed to be bi-Gaussian, with an input spot size and duration closely matching the experimental values. 

Figure \ref{fig:sim_vis} shows the transverse electron density and laser intensity profiles at three points in the channel for a conditioning pulse with a peak input intensity $I_\mathrm{peak} = \SI{6e17}{W.cm^{-2}}$. Close to the channel entrance, the far leading edge of the pulse ionizes the neutral gas surrounding the HOFI channel, creating a deep CHOFI channel in which the main body of the pulse is guided. The  position at which ionisation first occurs corresponds to the position at which $I(r, z)$ first exceeds $I_\mathrm{th}$ where $I_\mathrm{th}$, is the threshold laser intensity for ionization \cite{Shalloo:2018fy}.

Since the transverse intensity profile of the conditioning pulse is not perfectly matched to that of the lowest order mode of either the HOFI or CHOFI waveguides, the spot-size of the conditioning pulse oscillates by $\pm \SI{10}{\%}$ during the first few centimetres of propagation. However, this  variation primarily affects the wall thickness of the CHOFI channel, not its depth or matched spot size. Further, the higher-order modes excited by the conditioning pulse are attenuated with propagation, and the conditioning pulse adopts a stable configuration. This situation is shown in Fig.\  \ref{fig:sim_vis}(b-c), which shows that the radial extent of the  additional ionization is  reduced with propagation distance $z$. However, the high walls of the CHOFI channel remain, and are sufficiently thick, to ensure low-loss propagation of the bulk of the conditioning pulse. The conditioning pulse continues to propagate with low loss, and to generate a low-loss channel, until it can no longer ionize the neutral gas close to the shock front. Some temporal compression of the conditioning pulse is also evident, caused by leaky guiding of the leading edge in the HOFI channel and loss of energy to ionization, but this also has little effect on the generated CHOFI channel. We note that, close to the channel entrance, a plasma wakefield with a relative amplitude of  $\delta n_\mathrm{e} / n_\mathrm{e} = \SI{10}{\%}$ is driven by the conditioning pulse. This wakefield has no observable effect on the properties of the CHOFI channel, and in any case is expected to decay on a timescale of approximately \SI{10}{ps} \cite{jj_to_be} which is much shorter than the timescale observed in Fig.\ \ref{fig:exp_evolution} for significant evolution of the transverse electron density profile. To summarize, the CHOFI channel is relatively stable to variations in the properties of the conditioning pulse as it propagates, as well as to a degree of mis-matching to the HOFI channel. 

\begin{figure}[bt]
    \centering
    \includegraphics[width=80mm]{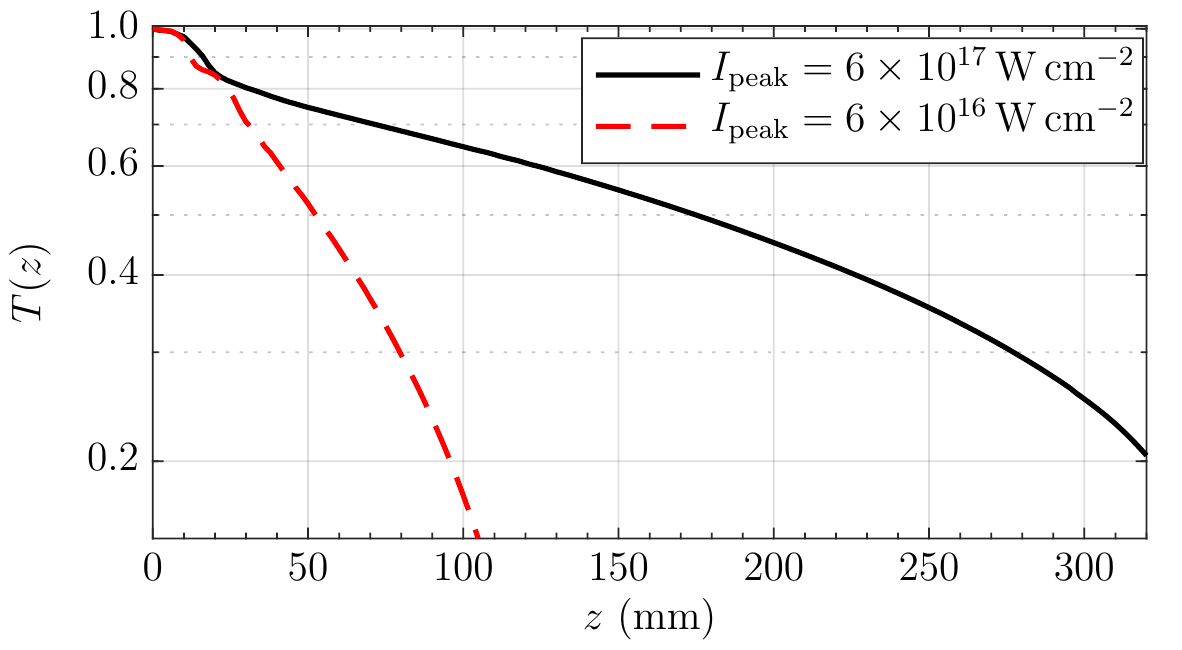}
    \caption{Calculated pulse energy transmission $T(z)$ of the \emph{conditioning} laser pulse as a function of propagation distance $z$, for peak input intensities of $\SI{6e16}{W.cm^{-2}}$ (dashed, red) and $\SI{6e17}{W.cm^{-2}}$ (solid, black).}
    \label{fig:sim_transmission}
\end{figure}

Figure  \ref{fig:sim_transmission} shows the pulse energy transmission $T(z)$ of the \emph{conditioning} pulse as a function of propagation distance $z$, for various input intensities. For intensities lower than $\sim \SI{5e15}{W.cm^{-2}}$, the conditioning pulse is not intense enough to ionize the neutral collar of gas, and hence it propagates in the HOFI channel and experiences high losses. For $I_\mathrm{peak} = \SI{6e16}{W.cm^{-2}}$, only the peak of the pulse is intense enough to ionize the neutral collar and generate a deep, thick channel. Thus, the leading edge etched away rapidly, and the length of the CHOFI channel is limited to $\approx \SI{100}{mm}$. In contrast, the highest intensity conditioning pulse shown in Fig.\  \ref{fig:sim_transmission} propagates with low losses over the length of the simulation. The calculated energy loss of the conditioning pulse is $\sim \SI{7}{mJ}$ per centimeter of waveguide. Note that the energy lost to ionization is estimated to be less than \SI{0.2}{mJ.cm^{-1}}, and hence the energy loss is dominated by etching of the leading edge in the leaky HOFI waveguide, and to driving a wakefield. 

The guiding properties of the CHOFI channels generated by the channel-forming and conditioning pulses were calculated by the numerical propagation code. First, the lowest-order mode was found by numerical propagation of an initially un-matched beam until the higher-order modes had been damped. Second, the power attenuation length was found by numerical propagation of this mode through the entire length of the CHOFI channel, including longitudinal variations in transverse profile of the CHOFI channel. For the conditions of Fig.\ \ref{fig:sim_vis}, the attenuation length was found to be $L_\mathrm{att} = \SI{2.5(1)}{m}$, which is nearly three orders of magnitude higher than that of the unconditioned HOFI channel.

\section{Discussion}
The experimental and simulation results above demonstrate that the leading edge of a conditioning laser pulse can ionize the neutral gas surrounding HOFI plasma channels to form a  conditioned HOFI channel which can guide the bulk of the conditioning pulse, and any subsequently injected laser pulse, with very low losses.  This mechanism is likely to have played a role in our recent experimental demonstration of guiding of high-intensity laser pulses in \SI{100}{mm} long channels \cite{Picksley:2020}. The simulations shown in Fig. \ref{fig:sim_vis} demonstrate the generation of CHOFI channels up to \SI{325}{mm} long, limited by erosion of the leading edge of the channel-forming pulse and pump depletion to the wakefield. We note that since the losses of the CHOFI channels are so low, it would be straightforward to extend the channel beyond this limit by employing a longer channel-forming pulse, or by using two or more channel-forming pulses.

Control over the axial density and matched spot size of CHOFI channels is in principle possible by adjusting the delay $\tau$ between the channel-forming and conditioning pulses, and the initial gas density. For example, increasing the delay $\tau$ from \SI{1.5}{ns} to \SI{4}{ns} reduces the axial density from $n_\mathrm{e} = \SI{2.4e17}{cm^{-3}}$ to $n_\mathrm{e} \approx \SI{7e16}{cm^{-3}}$ whilst maintaining an attenuation length $L_\mathrm{att} \approx \SI{2}{m}$. Since heating of the initial HOFI channel is independent of density, it is expected that it will be possible to create CHOFI channels at even lower densities, which, for example, would provide greater single-stage energy gain in a LWFA.

In addition, some control over the \emph{shape} of the CHOFI channel is possible by adjusting the delay $\Delta \tau$ between the conditioning pulse and the pulse to be guided. As shown in Fig.\ \ref{fig:exp_evolution}, for small values of $\Delta \tau$ the CHOFI channel has a core of approximately uniform electron density, surrounded by high and thick walls; this profile is a good approximation to the near-hollow plasma channel, which can provide independent control of the focusing and accelerating fields for electron and positron acceleration \cite{Schroeder:2013hr}. For larger values of $\Delta \tau$ the core of the channel develops an approximately parabolic profile.

An important feature of the work presented here is that the conditioning pulse is self-guided by ionization of the neutral gas at the edge of the initial HOFI channel. This self-guiding has three important consequences. First, as discussed below, the energy of the conditioning pulse is used efficiently, which reduces the total laser energy required to create the CHOFI channel. Second, the generated CHOFI channel is rather robust to variations of the parameters or pointing of the conditioning pulse. Thirdly, a wide range of channel parameters and shapes are accessible with a simple experimental setup. We note that the conditioning pulse does not require transverse shaping, and should have the same spot size and central wavelength as the main pulse to be guided; as such it could be generated very simply by introducing a small pre-pulse to the main pulse.

It is useful to estimate the total laser energy which would be required to generate a \SI{1}{m} long CHOFI channel. In our experiments approximately \SI{5}{mJ} per centimeter of HOFI channel was required for the channel-forming pulse. Simulations of the propagation of the conditioning pulse show that efficient self-guiding is achieved if it ionizes the neutral gas out to approximately $2w_\mathrm{m, CHOFI}$, which requires $I_{peak, cond} \gtrsim \SI{3.5e17}{W.cm^{-2}}$ for the $\SI{23}{\micro m}$ spot size used here. The calculations showed that when this is satisfied, the propagation losses are approximately \SI{7}{mJ.cm^{-1}}, from which we deduce that an input pulse energy of  $\SI{700}{mJ}$ is required for the conditioning pulse. Hence a total laser energy of \SI{1.2}{J} would be required, which is small compared to the energy required for the driving laser in a metre-scale LWFA, and almost an order of magnitude smaller than CHOFI channels generated by a high-order Bessel conditioning pulse \cite{howard_arxiv}.

The power attenuation lengths of the CHOFI channels demonstrated in this work are significantly larger than for other techniques capable of guiding pulses with peak intensities above $\SI{1E17}{W.cm^{-2}}$. A power attenuation length of $L_\mathrm{att} \approx \SI{0.15}{m}$ has been reported for capillary discharge waveguides \cite{Butler:2002zza}, and that for the  lowest-order mode of hollow capillary waveguides \cite{Cros:2002, Dorchies:1999vb} is $L_\mathrm{att} \approx \SI{0.9}{m}$ for a \SI{50}{\micro m} diameter capillary. Gonsalves et al.\ \cite{Gonsalves2020} have reported an energy transmission of \SI{85}{\%} for low-intensity pulses guided in \SI{90}{mm} long laser-assisted capillary discharges, corresponding to $L_\mathrm{att} \approx \SI{0.6}{m}$, assuming perfect coupling.

It is useful to estimate the electron energy gain of a plasma accelerator stage operating in the quasilinear regime, driven in the CHOFI channel shown in Fig.\  \ref{fig:sim_vis}. We note first that for these conditions $L_\mathrm{att} \gg L_\mathrm{pd}$, where $L_\mathrm{pd}$ is the pump depletion length of the wakefield driver. A drive laser pulse of energy \SI{1}{J}, and with a spot size matched to the CHOFI channel, has a normalized peak vector potential $a_0 = 1.13$. Guiding this over the dephasing length (\SI{0.1}{m}) would produce an energy gain of $\SI{2.8}{GeV}$. Higher energy gains could be achieved in channels with lower densities. For example, Leemans et al.\ \cite{Leemans:2010zz} have shown that a quasilinear \SI{10}{GeV} stage could be achieved with a \SI{40}{J} drive pulse ($a_0 = 1.4$) guided in a $\SI{0.6}{m}$ long plasma channel with an axial density of $n_\mathrm{e0} \approx \SI{1e17}{cm^{-3}}$ and a matched spot size of $\sim \SI{60}{\micro m}$. The work described in the present paper suggests that these challenging parameters could be achieved by CHOFI channels.

We note that very recently, the properties of the neutral gas collar surrounding a HOFI channel have been investigated by two-color interferometry \cite{hmm_self_guide}. In that work it was shown that a co-propagating (i.e. conditioning) pulse can form a low-loss (CHOFI) plasma channel by ionization of the gas collar, and guiding of a low-intensity pulse in the resulting CHOFI channel at $\Delta \tau \approx 0$ was demonstrated.

\section{Conclusions}
We have demonstrated through experiments and simulations that low-density, low-loss, metre-scale plasma channels can be generated by ionization of the neutral gas collar surrounding a hydrodynamic optical-field-ionized plasma channel. Channels with axial electron densities of $n_\mathrm{e0} \approx \SI{1E17}{cm^{-3}}$, a matched spot size of $w_\mathrm{m} \approx \SI{25}{\micro m}$, were generated experimentally. The channel depth was increased by a factor of ten compared to the unconditioned HOFI plasma channel and power attenuation lengths of up to $L_\mathrm{att} = \SI{26(2)}{m}$ were calculated from interferometric data.

Hydrodynamic simulations of the dynamics of the plasma formed by the channel-forming pulse show that a low-density HOFI plasma channel is created by the channel-forming pulse, and that this is surrounded by a collar of neutral gas. Particle-in-cell simulations show that the leading edge of a conditioning pulse injected into this structure ionizes the neutral gas in its transverse wings, to form a deep conditioned HOFI (CHOFI) channel which can guide the bulk of the conditioning pulse --- and any trailing laser pulses --- with very low propagation losses. 

The properties of CHOFI channels, including their shape, can be controlled by adjusting the initial gas density, and the delays between the channel-forming, conditioning, and guided pulse. Further, the channels are free-standing, which makes them immune to laser damage, and they can be generated with a total laser pulse energy of order \SI{1}{J} per metre of channel. These properties would seem to make them ideally suited to many applications in high-intensity light-matter interactions, including multi-GeV plasma accelerator stages operating at high pulse repetition rates. 

\section*{Acknowledgements}
The authors thank G. Hine for experimental assistance, and L. Feder, B. Miao and H. M. Milchberg for ongoing discussions. This work was supported by the UK Science and Technology Facilities Council (STFC UK) [grant numbers ST/P002048/1, ST/P002056/1, ST/N504233/1, ST/R505006/1]; the Engineering and Physical Sciences Research Council [studentship No.\ EP/N509711/1]; and the Central Laser Facility of the United Kingdom. This material is based upon work supported by the Air Force Office of Scientific Research under award number FA9550-18-1-7005. This work was supported by the European Union's Horizon 2020 research and innovation programme under grant agreement No. 653782. 

 \newcommand{\noop}[1]{}

\newpage

\section*{Supplemental Information}
\subsection*{Experimental set-up}
\begin{figure}[bt]
    \centering
    \includegraphics[width=84mm]{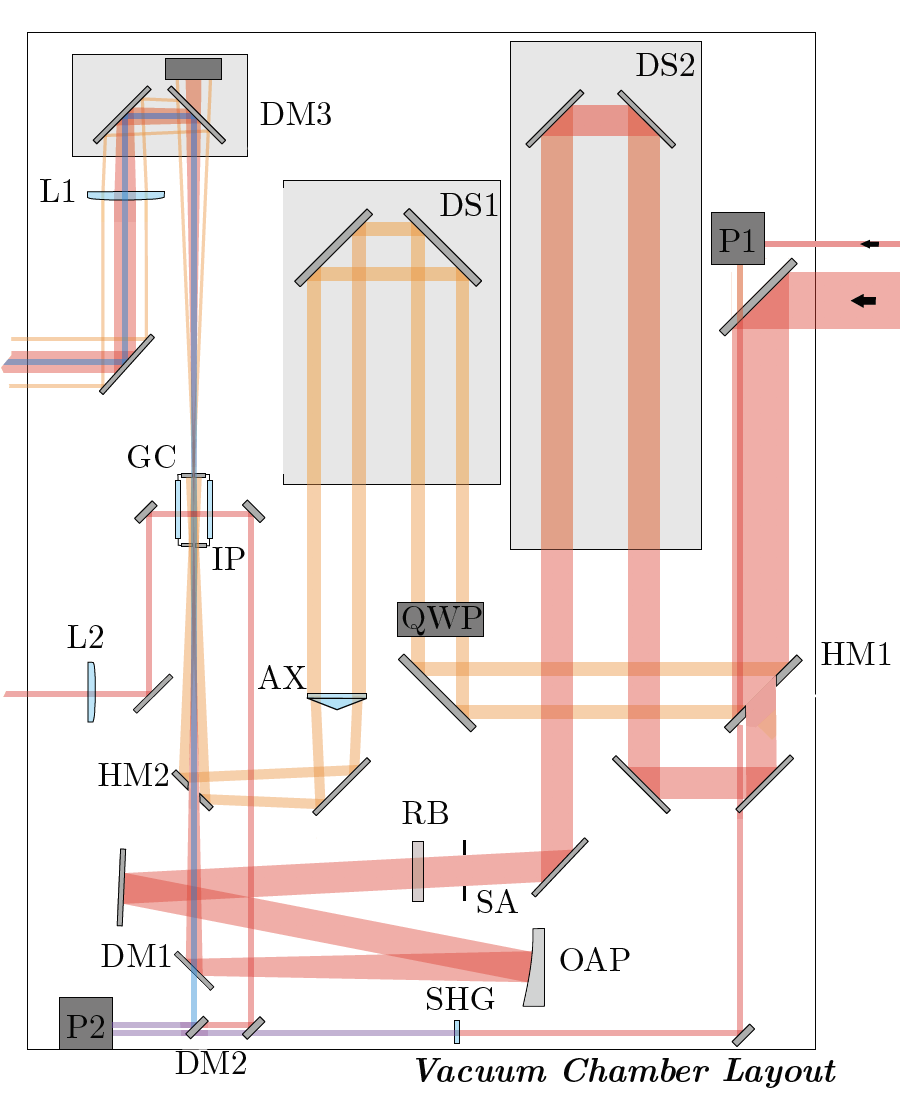}
    \caption{Detailed experimental layout showing vacuum chamber.}
    \label{fig:supp_setup}
\end{figure}

The experiment was carried out on the \SI{5}{Hz} repetition-rate TA2 Astra-Gemini Laser at the Central Laser Facilty, UK, providing a peak power of \SI{9.5}{TW}, using a setup very similar to the one described in Ref. \cite{Shalloo:2019hv} (see Fig. \ref{fig:supp_setup}). The compressed, collimated beam was approximately 55 mm in diameter. Upon entering the target chamber, the beam was split into the channel forming beam and the conditioning beam by a dielectric mirror (HM1) with a \SI{30}{mm} diameter hole drilled in the centre at $\SI{45}{\deg}$. The energy of both beams could be simultaneously varied by a half waveplate and polarizing chicane placed before the compressor.

The channel-forming beam was reflected from HM1 to form an annulus of light with collimated intensity $I^\mathrm{{coll}}_\mathrm{{ax}} \approx \SI{4.0e11}{W.cm^{-2}} $. This was reflected from a reterofelecting delay stage to allow for timing control, then focused by a fused silica axicon lens of base angle $\vartheta = \SI{5.6}{\deg}$. It was brought to the interaction point by a holed mirror (HM2) placed approximately \SI{120}{mm} from the gas cell entrance pinhole. The channel-forming beam had a peak axial intensity of $\SI{5e15}{W.cm^{-2}}$ inside the line focus, and a transverse profile that matched well to theory, $I(r) \propto J_0^2(\beta r)$ where $\beta = k[\arcsin(\eta \sin\vartheta) - \vartheta]$ and $k = 2\pi / \lambda_0$. The line focus extended throughout the length of the gas cell.

The conditioning pulse was transmitted through HM1 and passed through a retro-reflecting delay stage to allow the time between the channel-forming and conditioning pulses to be varied. It was then focused by an off-axis paraboloid (OAP) mirror of focal length $f = \SI{750}{mm}$ used at \textit{f}/25 to the entrance pinhole of the gas cell. The measured beam waist and Rayleigh range were $w_0 = \SI{22(3)}{\micro m}$ and $z_R = \SI{1.1(4)}{mm}$ respectively. By replacing the final turning mirror for an uncoated, optically flat wedge, the conditioning pulse could be operated at two intensities. The high-intensity pulse had peak intensity $I_\mathrm{{peak}}^\mathrm{{high}} = \SI{4.1e17}{W.cm^{-2}}$ at the entrance pinhole corresponding to $a_0 \approx 0.43$, whilst the lower intensity mode had a peak intensity $I_\mathrm{{peak}}^\mathrm{{low}} \approx \SI{1.6e16}{W.cm^{-2}}$.

The gas cell was constructed from aluminium and filled with Hydrogen gas through a \SI{5}{mm} inlet located \SI{10}{mm} from the entrance pinhole. Fused silica, optically flat windows were placed on each face to allow for transverse probing of the plasma channel. The entrance pinhole had a diameter of \SI{1.5}{mm}, and exit pinhole a diameter if \SI{800}{\micro m}. The distance between the two pinholes was measured to be \SI{16}{mm}. Pressure in the cell could be varied in the range \SIrange{5}{120}{mbar} and was measured using a pressure transducer located in a gas reservoir before the gas cell.

A separately compressed, $\SI{1}{mJ}$ probe pulse with the same central wavelength as the main pulse was used to probe the interaction transversely. The probe was passed transversely through the plasma at time $\tau$ after ionization by the channel-forming beam, which could be varied in the range \SIrange{0}{6}{ns} by a retroreflecting delay stage. Scattered light from the interaction region was collected and collimated by a \SI{50.2}{mm} diameter planoconvex lens of focal length $f_1 = \SI{250}{mm}$, and subsequently imaged onto a 12-bit CMOS detector by a second, planoconvex lens of diameter \SI{50.2}{mm} and $f_2 = \SI{1500}{mm}$. Before reaching the detector, this light was passed through a Mach-Zehnder interferometer to form equally spaced, straight fringes on the detector. The measured spatial resolution of the detector was $\SI{8.7(9)}{\micro m}$.

After the gas cell, light from the conditioning pulse was collected and collimated by a \SI{76.2}{mm} diameter, $f = \SI{500}{mm}$ focal length achromatic lens. A retro-reflecting stage placed before this lens allowed the object plane to be varied from the entrance to the exit pinhole of the gas cell. The collimated light was subsequently reimaged onto a 12-bit CMOS detector and an pyroelectric energy meter by $f = \SI{750}{mm}$, achromatic lens. Light from the channel-forming beam was blocked from reaching these diagnostics by an aperture.

\subsection*{Hydrodynamic Simulations}
\subsubsection*{Simulation parameters}
In order to study the gas and plasma densities prior to the interaction with the conditioning pulse, the channel expansion was reproduced by means of hydrodynamic simulations. The 2D simulations were performed using the Eulerian code FLASH 4.6.2\cite{fryxell2000flash}, capable of performing adaptive-mesh refinement inside the $\SI{100}{\micro m}\times \SI{100}{\micro m}$ simulation box. A 3-temperature model was employed in the calculations, which allowed for independent evolution of the electron, ion, and radiation species. Heat exchange was enabled to allow for energy exchange between electrons and ions. The energy diffusion and thermal conductivity modules were also activated, to improve the modelling of the species propagation.

Initially, the simulation box was filled with neutral atomic $H$ gas, at a pressure of \SI{50}{mbar} and a temperature of \SI{298}{K}. It should be noted that atomic $H$ was used due to the lack of modelling of the hydrodynamic properties of atomic $H_2$ available. The plasma column was initialised by configuring a region of higher temperature. As described in Ref.~\cite{Shalloo:2018fy}, the initial electron temperature profile for the plasma column created by our axicon can be described as a super-Gaussian, $T_{e}=T_{e,0}\cdot\exp\left(-(r/r_0)^{22}\right)$), where $T_{e,0} = $\SI{10}{eV} is the initial maximum temperature, and $r_0 = \SI{4}{\micro m}$ is the initial radius. Tabulated values were used for the equation of state (EoS) and ionization fraction of atomic $H$, with tables obtained from the commercially available PROPACEOS package~\cite{heliosCR}. Radiation transport was modelled using multi-group diffusion, with a total of 6 energy groups distributed between \SI{10}{meV} and \SI{12}{eV}. Tabulated values were used for the Rosseland and Planck opacities, also obtained from the PROPACEOS package.

\subsubsection*{Additional simulation results}

\begin{figure}[htb]
    \centering
    \includegraphics[width=\linewidth]{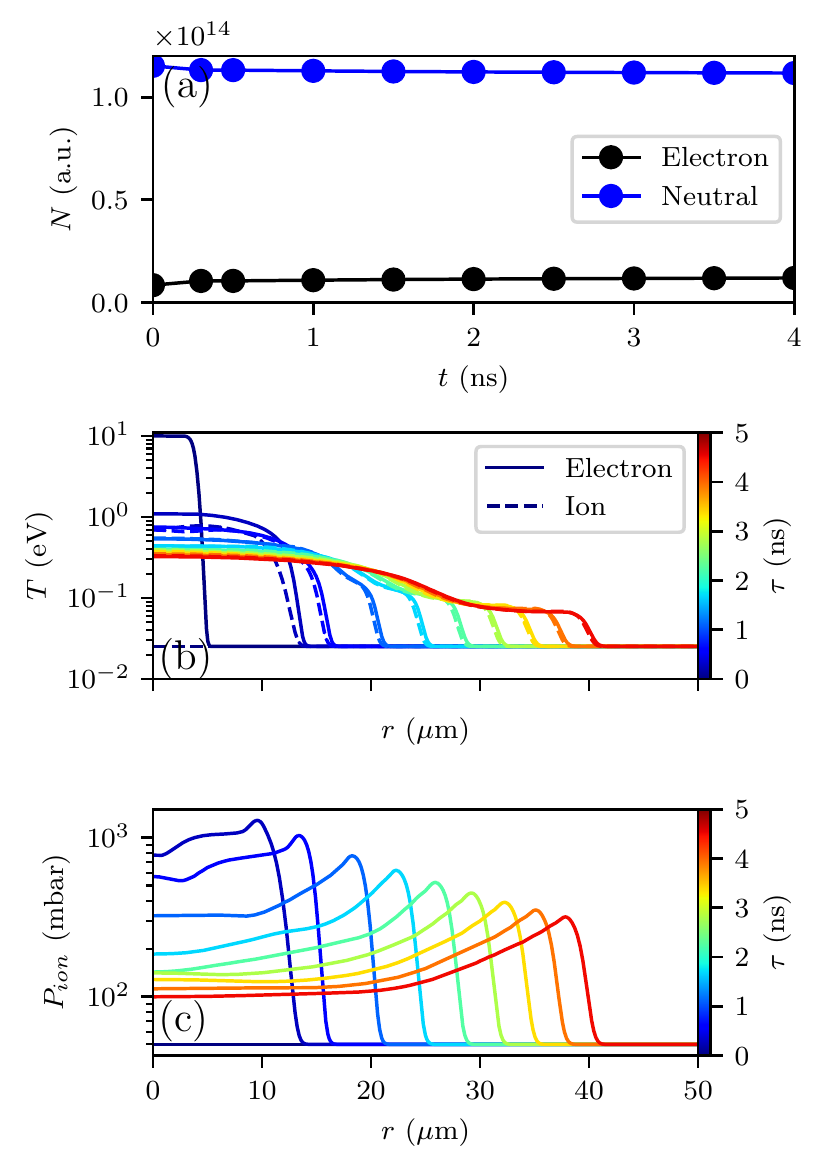}
    \caption{Additional results from the hydrodynamic simulations. (a) Temporal evolution of the number of free electrons (black) and neutral atoms (blue) throughout the plasma expansion. (b)  Temporal evolution of the electron ($T_e$, straight) and ion ($T_i$, dashed) temperature profiles at various delays $\tau$ given  by the color scale. (c) Temporal evolution of ion pressure profiles at various delays $\tau$ given  by the color scale.}
    \label{fig:flash_suppmat}
\end{figure}

The temporal evolution of the ionized and neutral components in the simulation was obtained by integrating the electron and neutral densities, respectively. As shown in Fig.~\ref{fig:flash_suppmat}(a), the fraction of each component remains constant throughout the expansion, in agreement with the experimental results in Fig.~3. This behaviour suggests that ionization dynamics is not the main cause for the accumulation of neutral gas.

Figure~\ref{fig:flash_suppmat}(b) shows the temporal evolution of the electron and ion temperature profiles. The temperature of both species exhibits an initial rapid evolution, until electron-ion thermalisation is reached at $\tau\simeq$\SI{1}{ns}. After this time, both species remain in equilibrium, and the evolution becomes significanly slower.

Finally, Figure~\ref{fig:flash_suppmat}(c) shows the temporal evolution of the ion pressure profiles. The ion pressure inner region of the channel is much greater than that beyond the shock front. Such increased pressure, together with the reduced ionization dynamics, indicate that the appearance of the neutral gas collar is caused by an snowplow-like effect.

\subsection*{Propagation simulations}

Simulations were initialized with an electron and neutral density profile equal to that simulated for $\tau = \SI{1.5}{ns}$. For the neutral species, the ionization energy was set to be \SI{13.6}{eV}, equal to that for a Hydrogen atom. The ionisation rates were calculated using the ADK model in the tunneling regime \cite{Lehe2016}. The laser pulses were assumed to have a Gaussian temporal profile with $\tau = \SI{40}{fs}$, and a Gaussian spatial profile with $w_0 = \SI{23}{\micro m}$, closely matching those in the experimental setup. 

%In EPOCH, ionization rates were calculated using the ADK model in the tunneling regime, and the Posthumus model \cite{Posthumus1997} in the BSI regime \cite{Arber2015, Lawrence-Douglas2013}. The threshold for ionization of Hydrogen in these regimes has previously been estimated to be $I_\mathrm{th} = \SI{2.5e14}{W.cm^{-2}}$ \cite{Shalloo:2018fy}. Collisional ionization was not included in the simulations. 

The simulation window was $400 \times 2100$ cells on the $(r, z)$ cylindrical grid, and co-propagated with the conditioning laser pulse. The grid resolution was \SI{0.046}{\micro m} and \SI{1}{\micro m} in the longitudinal and transverse directions respectively. Each cell was initialized with 4 particles per cell. At the transverse boundaries, perfectly matched layers were used to absorb outward traveling radiation from leaky modes. A linear density ramp of length \SI{100}{\micro m} was included at the start of the simulation to prevent unwanted reflections from a density discontinuity. The laser pulse was focused at the top of the ramp, though since the ramp length was much shorter than the Rayleigh length of the laser pulse, it did not affect laser coupling or propagation. 

The laser transmission was calculated by isolating the electric field of the laser from the plasma and integrating at each timestep
\begin{equation}
    T(z) = \frac{1}{\mathcal{E_\mathrm{init}}} \pi \epsilon_0 \int^{R_1}_{0} \int_{Z_{-}}^{Z_{+}} \left| E(r,z) \right|^2 \, r \ \mathrm{d}r \mathrm{d}z ,
\end{equation}
where $(0, R_1)$ and $(Z_{-}, Z_{+})$ are the edges of the simulation box in $r$ and $z$ respectively, and $\mathcal{E_\mathrm{init}}$ is the initial laser energy given by
\begin{equation}
    \mathcal{E_\mathrm{init}} = \left(\frac{\pi}{2} \right)^{3/2} \tau w_0^2 I_\mathrm{peak} .
\end{equation}

\end{document}